\documentclass[prl,twocolumn,showpacs,reprint]{revtex4-1}
\usepackage{amsfonts}
\usepackage{mathrsfs}
\usepackage{amsmath}
\usepackage{color}
\usepackage{graphicx}
\usepackage{bm}
\usepackage{amssymb}
\usepackage{xspace}
\usepackage{dcolumn}
\usepackage{longtable}
\usepackage{subfigure}
\usepackage{multirow}

\newcommand{\sgn}{\text{sgn}}
\makeatletter

\newcommand{\Rmnum}[1]{\expandafter\@slowromancap\romannumeral #1@}
\makeatother

\begin{document}
\title{Valley Chern Numbers and Boundary Modes in Gapped Bilayer Graphene}
\author{Fan Zhang$^1$}
\author{A. H. MacDonald$^2$}
\author{E. J. Mele$^1$}\email{mele@physics.upenn.edu}
\affiliation{
$^1$ Department of Physics and Astronomy, University of Pennsylvania, Philadelphia, PA 19104, USA\\
$^2$ Department of Physics, The University of Texas at Austin, Austin, TX 78712, USA
}
\begin{abstract}
Electronic states at domain walls in bilayer graphene are studied by analyzing their four and two band continuum models, by
performing numerical calculations on the lattice, and by using quantum geometric arguments. The continuum theories explain the
distinct electronic properties of boundary modes localized near domain walls formed by interlayer electric field reversal, by
interlayer stacking reversal, and by simultaneous reversal of both quantities. Boundary mode properties are related to
topological transitions and gap closures which occur in the bulk Hamiltonian parameter space. The important role played by
intervalley coupling effects not directly captured by the continuum model is addressed using lattice calculations for specific
domain wall structures.
\end{abstract}
 \maketitle

The electronic properties of few layer graphene systems depend sensitively on the atomic registry between neighboring layers
\cite{MM}. For bilayer graphene (BLG) with $AB$ stacking interlayer hybridization of orbitals on eclipsed lattice sites
gaps out half of the low energy degrees of freedom, replacing the pseudorelativistic description of single layer graphene
by a low energy theory in which quadratically dispersing chiral bands touch at discrete points in momentum
space \cite{McCann}. A perpendicular electric field further breaks inversion symmetry and creates a semiconductor in
which the gap size is determined by the electric field magnitude \cite{McCann,Ohta,Castro,ZhangABC} and saturated
at the strength of interlayer hybridization \cite{ZhangABC}. The possibility of exploiting this type of field tunable gap is
being vigorously pursued in ultra-clean dual gated devices \cite{Yacoby1,Yacoby2,Henriksen,Velasco,Baoetal}.

More recently it has been appreciated that the field-induced gap admits a topological interpretation \cite{Martinetal,SQH}.
The low energy theory for BLG can be represented by an effective two-band model
from which it is readily seen that inversion-symmetry breaking induces
large momentum-space Berry curvatures \cite{SQH,BerryRMP}.
The Berry curvatures have opposite sign near the two inequivalent Brillouin-zone
corners (valleys) at which gap is opened, so the
integral of the Berry curvature over the full Brillouin zone is zero.
Nonetheless, the integral of the Berry curvature within
a single valley is nonzero and this allows a topological analysis of the
valley-projected electronic spectrum. This idea has
been developed in a continuum analysis of the subgap electronic states bound to
a BLG domain formed by a sign reversal
of the electric field between the layers \cite{Martinetal,LiMorpurgo,NatPhys,Qiao,Jung}.
These electric-field walls (EFW) are predicted to bind
{\em pairs} of subgap chiral co-propagating boundary modes, an interesting feature that can be related to the change
in sign across a domain wall of a valley-projected topological index.

\begin{figure}[t!]
{\scalebox{0.42} {\includegraphics*{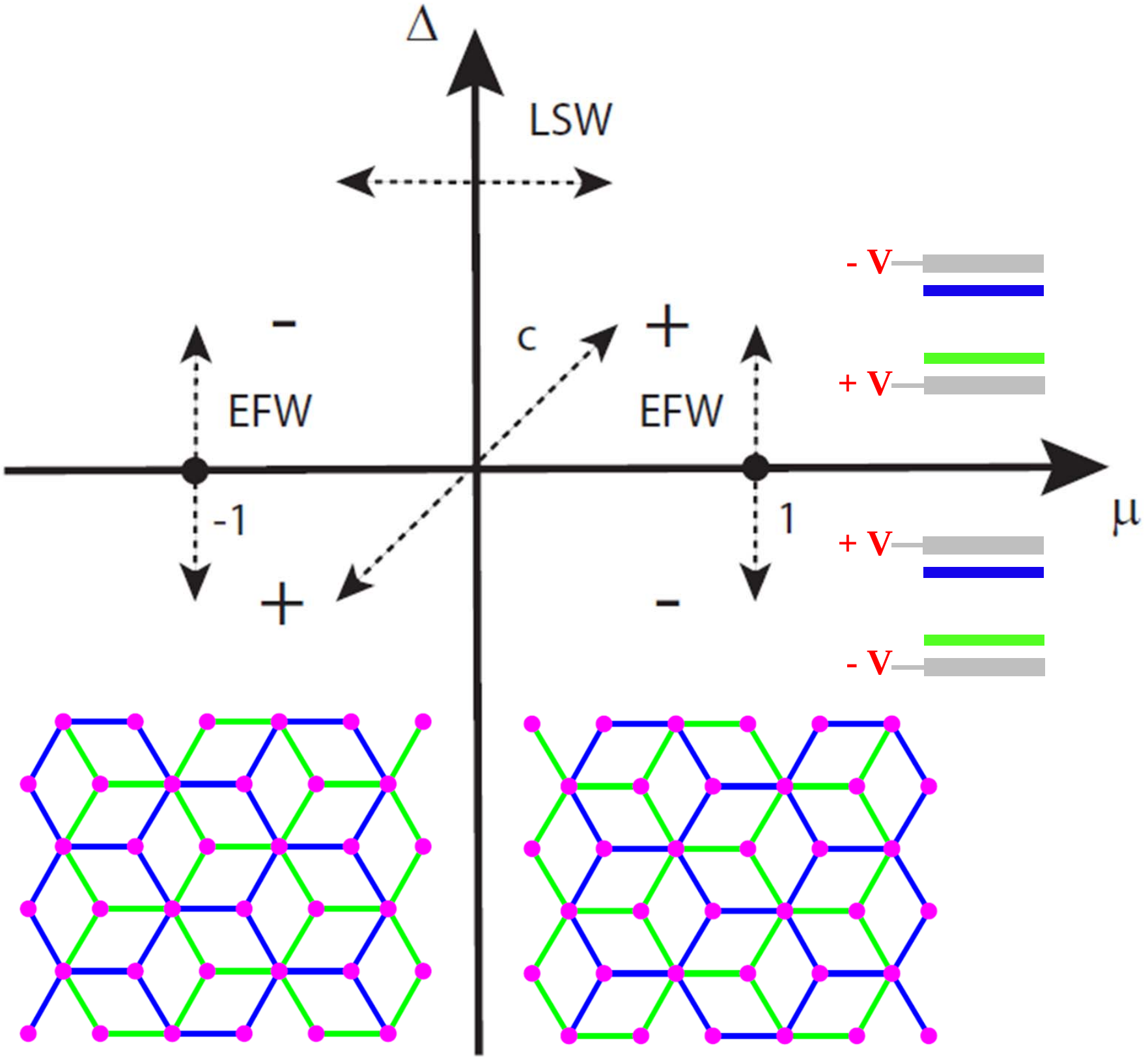}}}
\caption{\label{phasediagram} A phase diagram illustrating the distinct
valley-projected topological phases of BLG and the critical lines that separate them.
A sign change of $\Delta$ denotes a reversal of the interlayer electric field.
A sign change of $\mu$ denotes a transition from $AB$ to $BA$ interlayer registry.
The spectrum is gapped except along the lines $\Delta =0$ and $\mu=0$.}
\end{figure}

In this paper we examine a related BLG domain wall problem in which the interlayer electric field is {\em uniform} but the
layer stacking switches from $AB$ to $BA$. This new version of the problem changes the boundary conditions for matching the
electronic states of the two bounding phases and requires that we augment the two-band model of BLG
\cite{McCann,Martinetal,SQH} by accounting for all {\it four} sublattice degrees of freedom.  Nonetheless we find that
layer-stacking walls (LSW) bind electronic states with the same chiral structure as for the EFW studied previously. We make
this connection explicit by mapping the two problems onto each other within a family of four-band BLG Hamiltonians. Our results
demonstrate that the topological transition in a LSW structure is associated with a finite momentum gap closure in the
parameter space of four-band BLG Hamiltonians. We construct a phase diagram (Fig.~\ref{phasediagram}) which identifies the
different types of topologically protected states that appear in BLG samples in which both the interlayer electric field and
the layer stacking order vary in space. This analysis identifies yet a third type of domain wall in which the two pairs of
chiral modes within a single valley are coupled, gapping the spectrum and annihilating boundary modes. Our results are
supported by a continuum analysis of the domain wall states, lattice calculations for specific defect structures, and analysis
using quantum geometrical arguments. Taken together these elements provide a general framework for understanding the origin of
the valley-projected topological states in BLG, and their fragility in the presence of intervalley scattering.

The electronic states for BLG can be represented by four-component wavefunctions $\Psi=(\psi_{A_{T}},\psi_{B_{T}},\psi_{A_{B}},$ $\psi_{B_{B}})$ where $\psi$ denotes the atomic orbital centered on the $A$ or $B$ sites of the top or bottom layer.  At low energies
the Hamiltonian can be expanded for small ${\bm q}$ around the two inequivalent
Brillouin zone corners: ${\cal H}(\nu {\bm K} + {\bm q})$ with $\nu = \pm 1$ denoting $K$ and $K'$.  Using
Pauli matrices ${\bm \sigma}$ to represent operators that act on
the sublattice degree of freedom within a layer and ${\bm \tau}$ to represent
operators acting on the layer degree of freedom, the BLG Hamiltonian can separated into
layer diagonal and layer off-diagonal contributions by writing
${\cal H} = {\cal H}_0 + {\cal H}_{int}$.  We find that for
$AB$ stacked BLG in which the $A$ sites of top layer hybridize with the $B$ sites of the bottom layer,
${\cal H}_0 = \nu q_x \sigma_x \tau_0 + q_y \sigma_y \tau_0$ and ${\cal H}_{int} = \gamma (\sigma_x \tau_x - \sigma_y
\tau_y)/2$ where
energies are in units of $\hbar v$,
$\gamma = \gamma_1/\hbar v$, $\gamma_1$ is the near-neighbor
interlayer hopping amplitude, and $v$ is the electron velocity in an isolated layer.
For the reversed $BA$ stacking order the interlayer coupling term
becomes ${\cal H}_{int} = \gamma(\sigma_x \tau_x + \sigma_y \tau_y)/2$.
When present an electric potential difference $V$ adds  $\Delta\sigma_0 \tau_z$ with $\Delta = V/2 \hbar v_F$
to the Hamiltonian.

When $\gamma \gg \Delta$ it is convenient to eliminate the high energy degrees of freedom at $\pm \gamma$ to arrive at an
effective low energy two-band model \cite{McCann}
\begin{eqnarray}\label{H2}
\tilde{\cal H}_{\nu} = {\bm g}_{\nu} ({\bm q}) \cdot \tilde {\bm \sigma}\,,
\end{eqnarray}
where the $\tilde {\bm \sigma}$ matrices act on two component spinors $(\psi_{B_{T}},\psi_{A_{B}})$ in the low energy subspace for
$AB$ BLG and ${\bm g}_{\nu} ({\bm q})=(-(q_x^2 - q_y^2)/\gamma,2 \nu q_x q_y/\gamma,\Delta)$.
Eq.~(\ref{H2}) admits a geometrical interpretation in which the negative energy
eigenstates are spinors aligned with $-{\bm g}_\nu ({\bm q})$ and the filled band has a momentum space Berry curvature
\cite{SQH,BerryRMP}
\begin{eqnarray}\label{BerryCurv}
\Omega_{\nu} ({\bm q}) = \frac{-2\nu\gamma\,\Delta\,q^2}{(q^4 + \gamma^2 \Delta^2)^{3/2}}\,.
\end{eqnarray}
Because of the $\nu$ dependence in Eq.~(\ref{BerryCurv}) the integral of $\Omega_{\nu}({\bm q})$ over the full Brillouin zone
is zero and the filled valence band carries total Chern number $N=0$ as required by time reversal symmetry. However, for small
$\Delta$ the Berry curvature is strongly peaked at the gap minima near $K$ and $K'$. Consequently, the integral of
$\Omega_{\nu}({\bm q})$ over an individual valley is accurately defined and
the {\it valley Chern number} $N_{\nu}=-\nu\,\sgn(\Delta)=\pm 1$.
The valley Chern number changes by $\Delta N_{\nu}=\pm 2$ across an EFW
which can be associated with the appearance of {\it pairs} of valley-projected edge modes co-propagating along
the boundary. These chiral modes have been obtained by analytic solution
of the low energy two-band model in the presence of a
sharp EFW and by numerical solution for a spatially varying $\Delta({\bm r})$ that smoothly connects two electric-field
reversed states \cite{Martinetal}. As noted in previous work \cite{SQH,LiMorpurgo}, the introduction of a valley Chern number in this context is
approximate since strictly speaking the construction does not map the full periodic Brillouin zone onto the parameter space of
$\tilde {\cal H}_\nu$. Nonetheless, when $\Delta$ is small and intervalley scattering is absent the computed {\em change}
$\Delta N_\nu$ can be interpreted as a topological quantity, since $\Omega_{\nu}({\bm q})$
is integrated over a closed surface produced by ``gluing together" two integrals for the individual $N_\nu$ along a
common boundary.

\begin{figure}[t!]
{\scalebox{0.8} {\includegraphics*{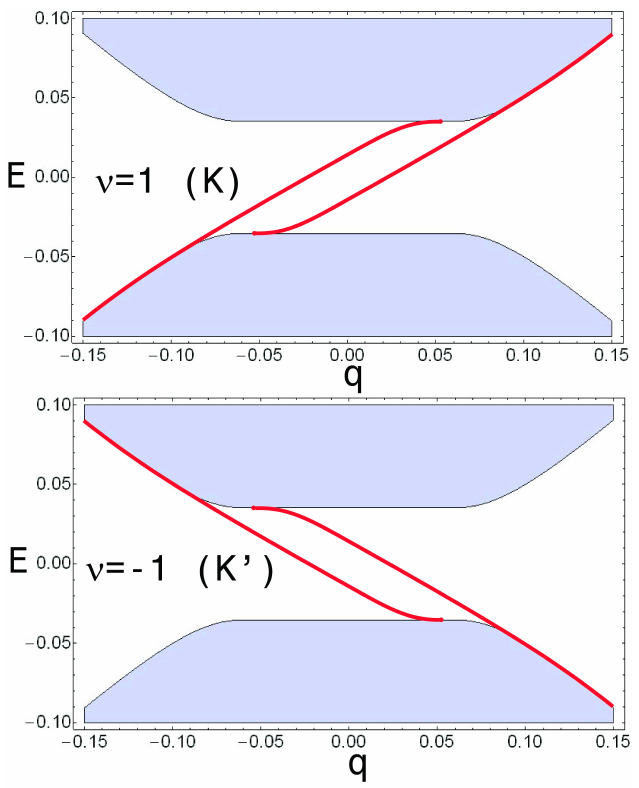}}}
\caption{\label{Valleys} Propagating chiral boundary modes (red) near the $K$ (top) and $K'$ (bottom) points
calculated using the four-band continuum model for a domain wall separating regions
with local $AB$ and $BA$ stacking registry
in the presence of a uniform layer-potential difference $\Delta$.
The shaded regions show the support of the continuous bulk spectrum
as a function of momentum ${\bm q}$ parallel to the domain wall.
The results were obtained for the parameter values $\Delta = 0.05, \, \gamma = 0.1$.}
\end{figure}

We now turn to the case of a LSW at which the bilayer registry reverses
from local $AB$ to local $BA$ with $\Delta$ held constant.
Crossing a LSW changes the interlayer coupling matrix ${\cal H}_{int}$ and switches
the orbitals that span its low energy subspace.
In this case evaluation of the Berry curvature requires
consideration of all {\em four} degrees of freedom in the bilayer Dirac problem.
Alternatively, one can identity the topological origin of LSW modes
by examining the residual phase twists induced
at large momentum $q$ in the eigenstates of the generalized Hamiltonian,
\begin{eqnarray}\label{HamLSW}
{\cal H}_{LSW} = \Delta \tau_z + \nu q_x \sigma_x + q_y \sigma_y + \frac{\gamma}{2} \left( \sigma_x \tau_x - \mu \sigma_y
\tau_y \right),
\end{eqnarray}
which reduces to the $AB\,(BA)$ forms when $\mu \to 1\,(-1)$.
For $q \gg |\Delta|,\gamma$ degenerate single layer states $\Psi_{\mu\nu}
({\bm q}) = (\psi_{\mu\nu,T},\psi_{\mu\nu,B})$ deep in the filled band with energies $E=-|{\bm q}|$ are split by $\Delta$ and are mixed by
$\gamma$ in the projected Hamiltonian
\begin{eqnarray}\label{Hamval}
{\cal H}_{\nu}^{-} = -|q| \lambda_0 + \Delta \lambda_z - \frac{\nu \gamma}{4} \left(e^{i \mu \nu \phi} \lambda_+ + e^{-i \mu
\nu \phi} \lambda_- \right)\,,
\end{eqnarray}
where ${\bm \lambda}$ are $2 \times 2$ Pauli matrices acting in the
$\Psi_{\mu\nu}$ subspace and $\phi = \arctan(q_y/q_x)$. For large $q$ the eigenstates $\Psi_{\mu\nu,\pm}$ written in the
original four orbital basis are
\begin{eqnarray}\label{wfs}
\Psi_{\mu \nu,\pm} &=& \frac{e^{i \alpha_{\mu \nu,\pm} (\phi)}}{2}\left(%
\begin{array}{l}
\sqrt{1 \pm \frac{\Delta}{\xi}}  \\
\nu \sqrt{1 \pm \frac{\Delta}{\xi}} \, e^{i \nu \phi}  \\
\mp \nu \sqrt{1 \mp \frac{\Delta}{\xi}} \, e^{-i \mu  \nu \phi} \\
\mp  \sqrt{1 \mp \frac{\Delta}{\xi}} \, e^{i (1- \mu) \nu \phi}
\end{array}%
\right)\,,
\end{eqnarray}
where $\xi = \sqrt{\gamma^2/4 + \Delta^2}$  and we explicitly display the overall $U(1)$ phases $\alpha_{\mu \nu,\pm}$.
Using Eq.~(\ref{wfs}) we calculate the momentum space Berry connection
\begin{eqnarray}
{\cal A}_{\mu \nu,\pm} = {\rm Im} \, \langle \psi_\pm | \partial_\phi  \psi_\pm \rangle
  =  \frac{(1-\mu)\nu}{2} \pm \frac{ \mu \nu \Delta}{2 \xi}  + \frac{\partial \alpha_{\mu \nu, \pm}}{\partial \phi}\,.
  \nonumber\\
\end{eqnarray}
The change in the valley Chern numbers upon passing from the $\mu = 1$ to $\mu=-1$ states
is obtained from  the loop integral of the
trace of $-\mu{\cal A}_{\mu\nu,\pm}$ over $\mu$ and band indices $\pm$, which reads
\begin{eqnarray}\label{DeltaN}
\Delta N_\nu = 2\,\nu + m_{-1,\nu,+} +m_{-1,\nu,-} - m_{1,\nu,+} - m_{1,\nu,-}\,.
\end{eqnarray}
Here $m_{\mu\nu,\pm}$ are integer valued winding numbers of the overall phases $\alpha_{\mu \nu,\pm}$. The $\Delta$ dependence of this result
vanishes after tracing over the filled bands, demonstrating that the valley Chern number in BLG is shared among all the
occupied bands rather than being confined just to its low energy states as is often assumed.  $\Delta N_\nu$ is a
topological index provided that the difference is evaluated in the same gauge for the two bounding phases, which requires that
$m_{-1,\nu,\pm} = m_{1,\nu,\pm}$ for this boundary.   It follows that
$\Delta N_\nu = 2\,\nu$ and therefore that a domain wall
separating insulating regions with local $AB$ and $BA$ registry
will also confine pairs of valley-projected chiral modes propagating
along the boundary with opposite velocities in the two valleys.
Fig.~\ref{Valleys} confirms this result by showing the spectra calculated
by matching the full four component wavefunctions of Eq.~(\ref{HamLSW}) across a sharp boundary where $\mu$
switches from $1$ to $-1$.

\begin{figure}[t!]
{\scalebox{0.175} {\includegraphics*{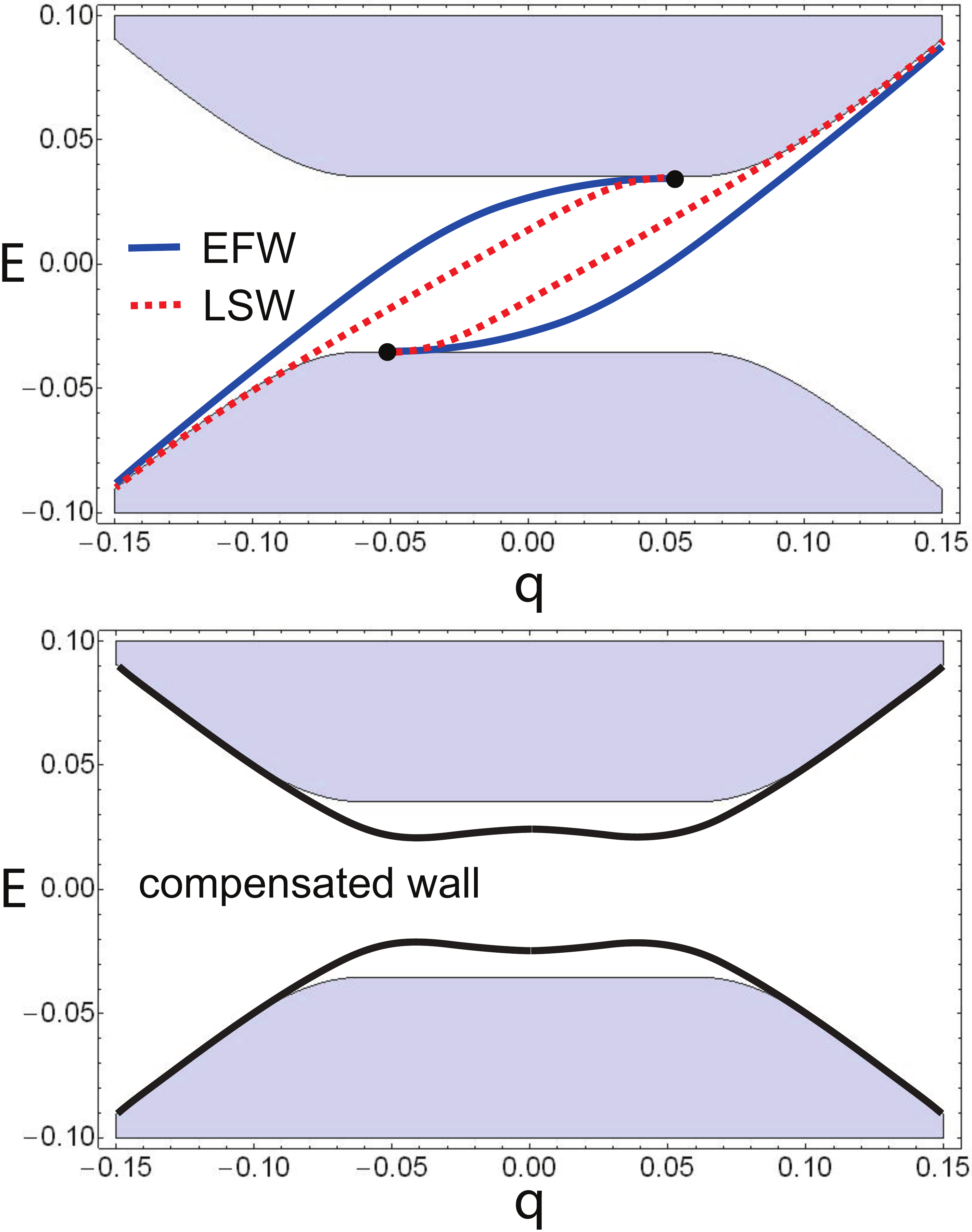}}}
\caption{\label{Compared} Top: Comparison of valley $K$ domain wall spectra for
electric-field and layer-stacking walls.  Both walls support a pair of co-propagating chiral modes.
Bottom: Interface spectrum calculated with a four-band theory for a topologically compensated boundary
at which both layer stacking and electric field change sign.  In this case
the boundary spectrum is completely gapped even in a continuum model.}
\end{figure}

Using Eq.~(\ref{wfs}) we find that at large $q$ the wavefunctions on the two sides of the LSW are related by a gauge
transformation,
\begin{eqnarray}\label{wfs3}
\Psi_{\mu,\pm} \mapsto e^{ i \mu \nu (1 - \tau_z) \phi } \Psi_{\mu,\pm}\, ,
\end{eqnarray}
with a different phase twist induced in each layer.  It follows from the accumulation of
internal phases in the two layers that $\Delta N_\nu=2\mu\nu$.
EFW walls, where the sign of the potential difference between layers $\Delta$ switches but the atomic registry
$\mu$ does not change, can be analyzed similarly.
Although $\Delta$ does not appear explicitly in $\Delta N_{\nu}$ in
Eq.~(\ref{DeltaN}) there is an implicit dependence through the $U(1)$ phase prefactors.  We find
\begin{eqnarray}
{\cal H}_{LSW} (-\Delta,\mu)= \tau_x \,{\cal H}_{LSW}(\Delta,-\mu)\, \tau_x\,,
\end{eqnarray}
{\it i.e.} that a sign reversal in $\Delta$ can be absorbed in a sign change of $\mu$ combined with a change of basis.
Using this construction the negative energy eigenstates at large ${\bm q}$ on the two sides
of the EFW are related by the following $U(1)$ gauge transformation
\begin{eqnarray}\label{wfs2}
\Psi_{\mu,\pm} \mapsto \mp \nu e^{i \mu \nu \phi} \Psi_{\mu,\pm}\,,
\end{eqnarray}
which produces the same overall $\Delta N_\nu = 2 \mu \nu$ for the EFW. Eqs.~($\ref{wfs3}$)-($\ref{wfs2}$) compactly express the relation between these two
different types of domain wall in the four-band theory.
This is also illustrated in Fig.~\ref{Compared} which compares the
valley $K$ spectra computed for an EFW and a LSW showing their common chiral boundary modes.

In the LSW case, unlike the EFW case,
analyzing the continuity of wavefunctions across the interface requires consideration of all four
bands.  The common topological origin of the domain wall spectra therefore
becomes apparent only in a four-band continuum theory.
Nevertheless, by integrating out the high energy bands at energies $\sim \pm \gamma$  we are able to construct a two-band
effective model away from the domain wall in which for either case $\Delta N_\nu$ is assigned to a sign change of the
Berry curvature of the lower band. In this approach, Eq.~(\ref{H2})
reads ${\bm g}_{\nu} ({\bm q})=(-(q_x^2 -
q_y^2)/\gamma,2\mu\nu q_x q_y/\gamma,\Delta)$ with $\tilde {\bm \sigma}$ {\it layer} Pauli matrices that act on
different spinors in the $\mu = \pm 1$ cases: on $(\psi_{B_T},\psi_{A_B})$ for $\mu=1$
and on $(\psi_{A_T},\psi_{B_B})$ for $\mu=-1$. Because of
inversion symmetry breaking, the valence band acquires a momentum space Berry curvature \cite{SQH,BerryRMP},
\begin{eqnarray}\label{BC}
\Omega_{\nu} ({\bm q}) = \frac{-2\mu\nu\gamma\,\Delta\,q^2}{(q^4 + \gamma^2 \Delta^2)^{3/2}}\,,
\end{eqnarray}
which integrates over a single valley to $-\mu\nu\sgn(\Delta)$.
Obviously, the valley Chern number changes by two across either a EFW or a LSW. Based on the bulk-boundary correspondence,
pairs of valley-projected edge modes should co-propagate along the interface.

\begin{figure}
{\scalebox{0.65} {\includegraphics*{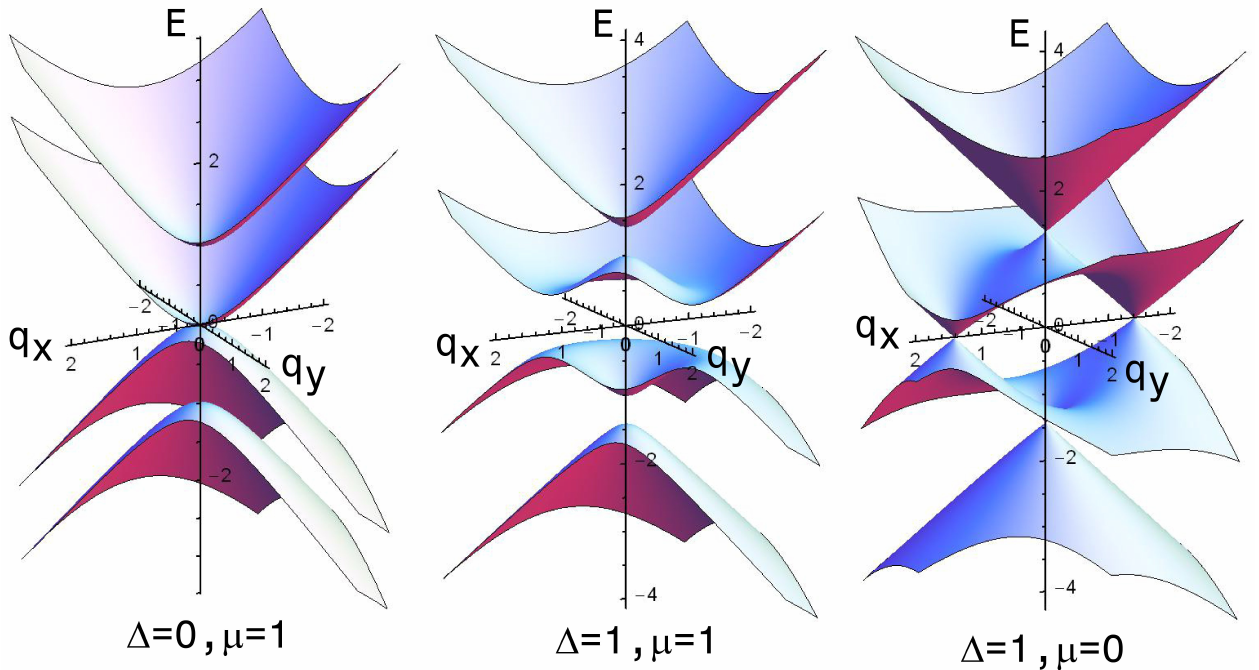}}}
\caption{\label{ThreeBands} Bulk band structures of three forms of BLG. Left: BLG with uniform AB registry
and no interlayer electric field characterized by quadratic touching between its two low energy bands.
Middle: the field-induced gapped phase in which the BLG degeneracy is lifted by an interlayer electric field.
Right: a critical state with nonzero interlayer electric field where gap closure occurs at $\mu=0$ as BLG
crosses from $AB$ to $BA$ registry.
The zero energy gap closures in the left and right panels separate four distinct gapped phases
of BLG as shown in Fig.~\ref{phasediagram}.}
\end{figure}

The $\Delta-\mu$ plane phase diagram in Fig.~\ref{phasediagram} identifies distinct BLG topological phases.  Phase boundaries
occur along the $\mu$ and $\Delta$ axes where the spectrum of ${\cal H}_{LSW} (\Delta,\mu)$ undergoes gap closures at $E=0$.
${\cal H}_{LSW}(\Delta=0,\mu=1)$ describes an ungapped BLG system in which quadratic band crossing occurs exactly at ${\bm
q}=0$, as seen in the left panel of Fig.~\ref{ThreeBands}. The gaps that open for the case of $\Delta \neq 0$ and $\mu=\pm 1$
are the electric field induced gaps easily understood within a two-band model. The boundary with $\Delta \neq 0$ and $\mu=0$
also has a gap closure, but it occurs at two finite momenta $q_x = \pm \sqrt{\Delta^2+\gamma^2/4}$ along the $q_y=0$ line where
band crossing is possible because $\sigma_x$ is a constant of the motion. For deviations in either $q_x$  or $q_y$ the
degeneracies at the band-crossing points are lifted at linear order, implying the conical gap closure illustrated in the right
panel of Fig.~\ref{ThreeBands}. When $\mu \neq \pm 1$ the original quadratic gap closure fissions into a pair of linear Dirac
singularities each of which carries half the original winding number. Trajectories in the Hamiltonian parameter space that
connect these topologically distinct ground states and involve different parameter values can shift the momenta at which the
gap closures occur, but cannot eliminate them.  For example, when $\Delta \neq 0$ but $\gamma =0$ in Eq.~(\ref{HamLSW}) the
layers decouple and the gap closure at $E=0$ degenerates to a closed Fermi ring with radius $\Delta$.

Fig.~\ref{phasediagram} also illustrates the possibility of a third type of compensated domain structure (labeled c) at which
both the layer registry and interlayer electric field are reversed.  Variation of local band parameters along
this line connects two bilayer states that are distinct but have $\Delta N_\nu=0$.
As illustrated in the the lower panel of Fig.~\ref{Compared}, spectra obtained by
matching solutions across this compensated domain wall
demonstrate that it hosts two pairs {\it counter-propagating modes
within the same valley}, which hybridize and completely gap the spectrum.

The continuum model is able to explain the topological origin of the gapless interface modes. However, the short-range physics
near the domain wall, which may be of essential significance, is not captured in the continuum Hamiltonian.
Importantly, the single valley physics that protects the chiral domain wall solutions can be preempted by sufficiently strong
large momentum scattering that acts to recouple states in the two valleys. In fact, Fig.~\ref{Valleys} suggests that these
single valley domain wall modes ultimately reconnect with each other.
To study this further we construct a specific lattice
model and use it to investigate how both lattice and interfacial effects,which couple the two valleys,
influence the domain wall modes.

\begin{figure}[b!]
{\scalebox{0.55} {\includegraphics*{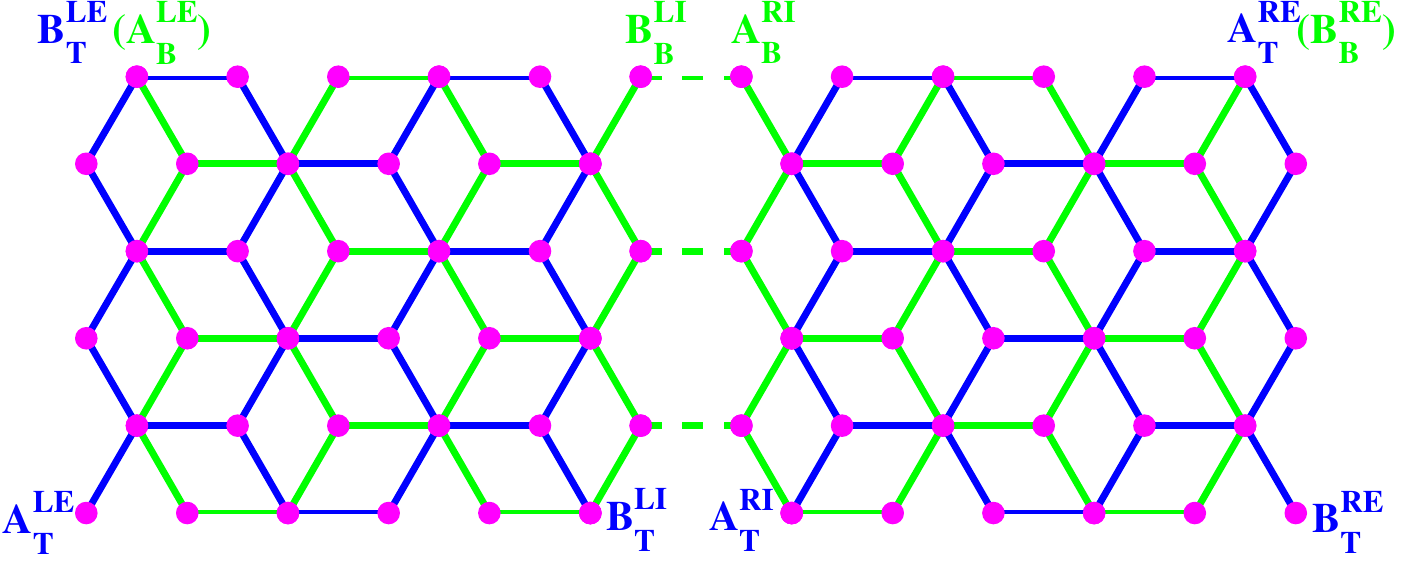}}}
\caption{\label{gb} The simplest LSW separating BLG into a left (L) domain with $B_{T} A_{B}$ stacking ($\mu=-1$)
and a right (R) domain with $A_{T} B_{B}$ stacking ($\mu=1$).  When the BLG is uniformly gapped,
gapless modes emerge along the outer zigzag edges (E) as well as along the LSW interfaces (I).
The lattices are continuous in the bottom (B) layer but have a straight crack in the top (T) layer.
The dashed lines denote the tunneling between the domains within the continuous layer.}
\end{figure}

\begin{figure}[t!]
{\scalebox{0.54} {\includegraphics*{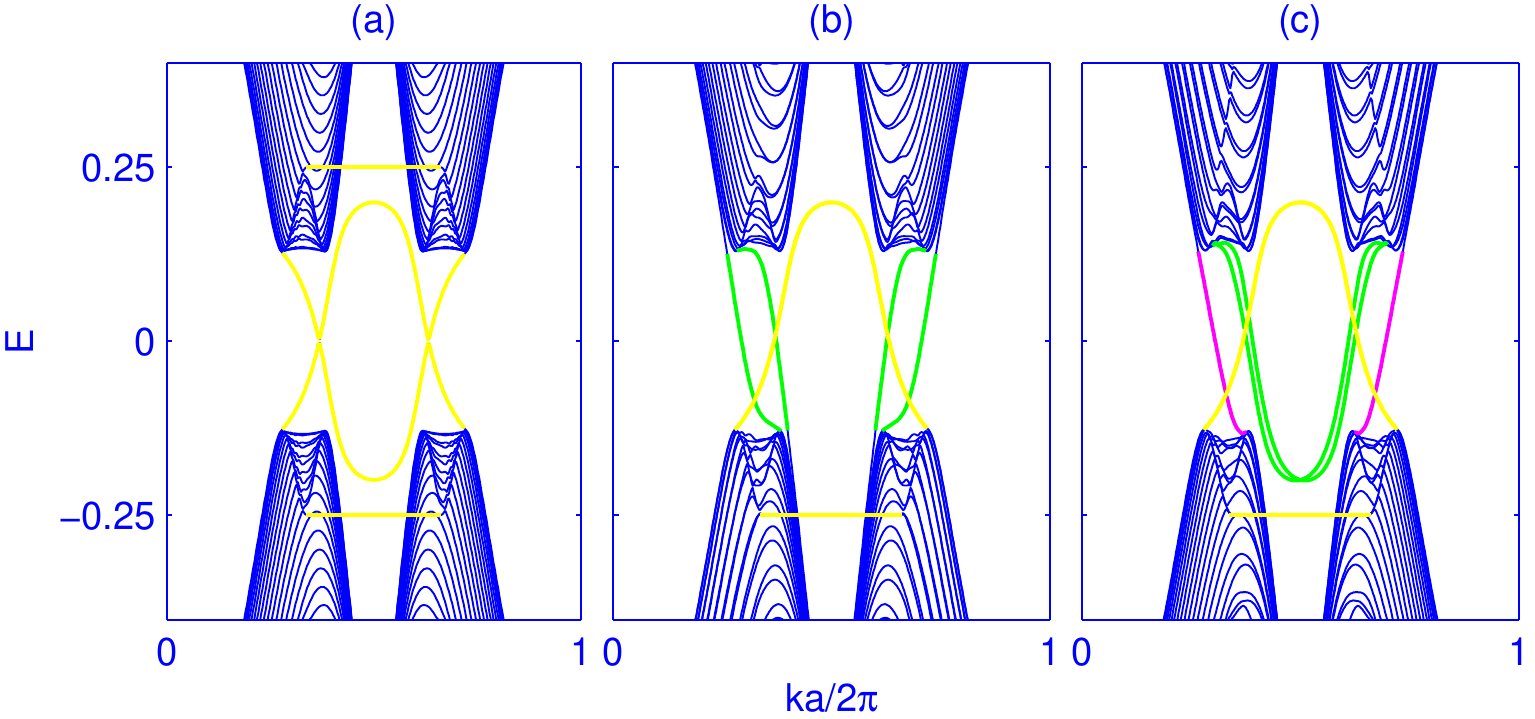}}}
\caption{\label{fg6} Gapless modes for (a) uniform gapped BLG, for (b) gapped BLG with an EFW,
and for (c) gapped BLG with a LSW as depicted in Fig.~\ref{gb}.
The yellow states localize on the outer zigzag boundary and they are doubly degenerate in (b) and (c).
The green states localize on the EFW in (b) and on the LSW in (c).
In (c) the green (magenta) LSW states localize on the broken (continuous) layer.
To illustrate these different cases, we choose parameter values: $t=1$, $\gamma_1=0.3$, and $V/2=0.25$.}
\end{figure}

As depicted in Fig.~\ref{gb}, we consider the simplest LSW, {\em i.e.}, a grain boundary separating BLG into left and right
domains. Near the LSW, the lattices are continuous in one layer but fractured along a zigzag edge in the other. This introduces
additional zigzag boundaries in the broken layer and allows switching of the bulk stacking order from $B_{T}
A_{B}$ ($\mu=-1$) on the left to $A_{T} B_{B}$ ($\mu=1$) on the right. For comparison, we first
calculate the band structures for the case of uniform gapped BLG
and for the case of gapped BLG with an EFW at which stacking order
is preserved.  As expected and shown in Fig.~\ref{fg6}(a) and (b), quantum valley Hall edge
states \cite{SQH} and two flat bands appear at the outer zigzag edges in uniformly gapped BLG.
In the sample with an EFW there is an additional pair of co-propagating chiral gapless modes
which emerge at each valley.  Fig.~\ref{fg6}(c) shows the situation for a
LSW with a uniform interlayer electric field; surprisingly there are three instead of two gapless modes per valley in this case.

\begin{figure}[b!]
{\scalebox{0.54} {\includegraphics*{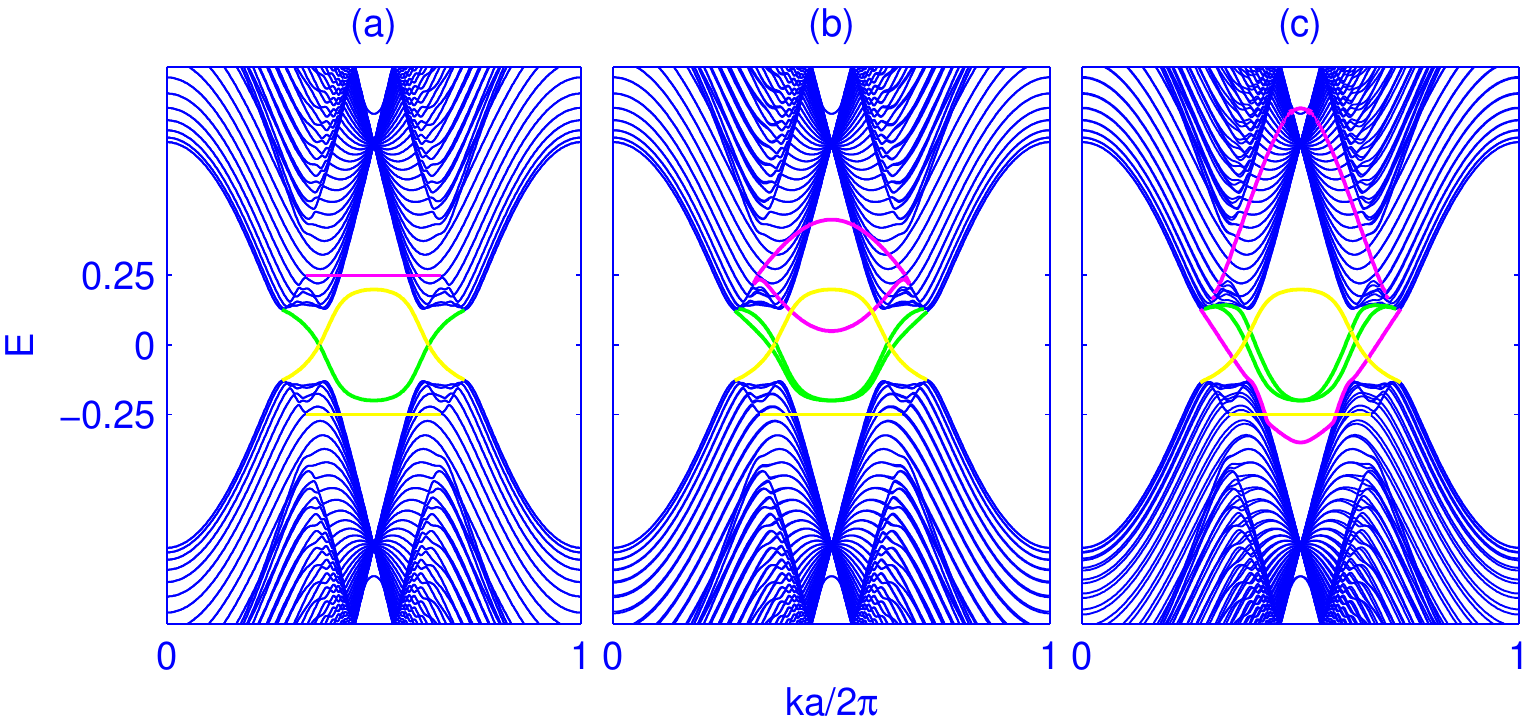}}}
\caption{\label{fg7} Evolution of gapless modes for gapped BLG as a function of
hybridization across the LSW. The tunneling parameter $t_{c}$ is
respectively $0$ in (a), $0.2$ in (b), $0.6$ in (c), and $1$ in Fig.~\ref{fg6}(c).
The other parameters have the same values as in Fig.~\ref{fg6}.
In panel (a) all the yellow, green, and magenta states are doubly degenerate.}
\end{figure}

We investigate this problem further by studying the dependence on the tunneling amplitude $t_{c}$ across the LSW shown as the dashed lines in
Fig.~\ref{gb}. Without tunneling (Fig.~\ref{fg7}(a)), the
boundary mode spectrum yields two copies of the gapped BLG spectrum shown in Fig.~\ref{fg6}(a), and thus there are two chiral
gapless modes in each valley as anticipated by the continuum model.
The flat bands represent the states localized on the grain
boundary lines $B_{B}^{LI}$ and $A_{B}^{RI}$ in Fig.~\ref{gb}. The leading effect of turning on the tunneling
is that the pair of degenerate flat bands (magenta bands in Fig.~\ref{fg7}) are split and become dispersive,
as described in Fig.~\ref{fg7}(b) and (c).
When the tunneling is larger than the electric field induced gap, the flat band split downward is pushed down
to the valence band and becomes the third gapless mode shown in Fig.~\ref{fg6}(c).

The other two gapless modes (green bands in Fig.~\ref{fg7}) localized on the grain boundary of the broken layer are almost degenerate due to the inversion symmetry between the lines $B_{T}^{LI}$ and $A_{T}^{RI}$ in Fig.~\ref{gb}. This degeneracy is exact at $ka=\pi$
and can be lifted by breaking the inversion symmetry between the left and right domains.
We further find that a local potential
$U_{loc}$ on $B_{T}^{LI}$ or $A_{T}^{RI}$ can raise and lower the energies of the green bands.
Similarly, a line potential on $B_{B}^{LI}$ or $A_{B}^{RI}$ can change the energies of the magenta bands.
In view of these results we propose a criterion controlled by a hierarchy of energy scales
to determine the number of fragile gapless modes in the atomically abrupt LSW shown in Fig.~\ref{gb}:
\begin{eqnarray}\label{eqa}
U_{loc}-\frac{V}{2} &<& -\frac{\Delta_{gap}}{2}\,,\\
\label{eqb}
U_{loc}+\frac{V}{2}\pm t_{c}&<& -\frac{\Delta_{gap}}{2} \,,
\end{eqnarray}
where $\Delta_{gap}/2$ is half size of the field-induced gap which saturates if $V$ exceeds a critical value \cite{ZhangABC}.
Two gapless channels emerge if Eq.~(\ref{eqa}) is satisfied, but extra gapless channels can
also appear from the flat bands if Eq.~(\ref{eqb}) is fulfilled.

When the LSW is made smooth in the sense that it does not
produce sufficiently strong intervalley coupling,
the tunneling amplitudes near the domain wall in both layers are almost the same as the pristine ones.
In such a case, the tunneling between the left and right domains in the broken (continuous) layer
would strongly split the two green (magenta) bands at $ka=\pi$.
As a result, only one green and one magenta bands in Fig.~\ref{fg7} survive in the band gap,
recovering our earlier continuum results.

We conclude that the gapless interface modes at a LSW are topologically stable only if the potential difference between layers
is the dominant energy scale, so that valley is approximately a good quantum number. In the general case the number of domain
wall modes can be any integer from $0$ to $4$ depending on the criteria like that implied by Eq.~(\ref{eqa}) and (\ref{eqb}).
The valley-projected topological-state physics of BLG is illustrative of similar physics which occurs in all multi-layer
graphene systems \cite{ZhangABC,SQH} and is sensitive to stacking order, and to perpendicular electric fields.

{\em Note added.---}
After the finalization of this work, a complementary preprint \cite{Kim}, which covers closely related material, has appeared.

This work is supported by DARPA under grant SPAWAR N66001-11-1-4110, by the Department of Energy, Office of Basic Energy Sciences under contract DE-FG02-ER45118, and by the Welch Foundation grant TBF1473.


\begin{thebibliography}{11}

\bibitem{MM} H. Min, A. H. MacDonald, Phys. Rev. B {\bf 77}, 155416 (2008).

\bibitem{McCann} E. McCann and V. I. Fal'ko, Phys. Rev. Lett {\bf 96}, 086805 (2006).

\bibitem{Ohta} T. Ohta, A. Bostwick, T. Seyller, K. Horn, and E. Rotenberg, Science {\bf 313}, 951 (2006).

\bibitem{Castro}  E. V. Castro, K. S. Novoselov, S. V. Morozov, N. M. R. Peres, J. M. B. Lopes dos Santos, J. Nilsson,
F. Guinea, A. K. Geim and A. H. Castro Neto, Phys. Rev. Lett. {\bf 99}, 216802 (2007).

\bibitem{ZhangABC} F. Zhang, B. Sahu, H. Min and A. H. MacDonald, Phys. Rev. B {\bf 82}, 035409 (2010).

\bibitem{Yacoby1} B. E. Feldman, J. Martin and A. Yacoby, Nature Physics {\bf 5}, 889 (2009).

\bibitem{Yacoby2} J. Martin, B. E. Feldman, R. T. Weitz, M. T. Allen and A. Yacoby, Phys. Rev. Lett. {\bf 105}, 256806 (2010).

\bibitem{Henriksen} E. A. Henriksen and J. P. Eisenstein, Phys. Rev. B {\bf 82}, 041412(R), (2010).

\bibitem{Velasco} J. Velasco, L.Jing, W. Bao, Y. Lee, P. Kratz, V. Aji, M. Bockrath, C. N. Lau, C. Varma, R. Stillwell,
D. Smirnov, F. Zhang, J. Jung, and A. H. MacDonald, Nature Nano. {\bf 7}, 156 (2012).

\bibitem{Baoetal} W. Bao, J. Velasco, F. Zhang, L. Jing, B. Standley, D. Smirnov, M. Bockrath,
A. H. MacDonald, and C. N. Lau, Proc. Nat. Acad. Sci. {\bf 109}, 10802 (2012).

\bibitem{Martinetal} I. Martin, Y. M. Blanter, and A. F. Morpurgo, Phys. Rev. Lett. {\bf 100}, 036804 (2008).

\bibitem{SQH} F. Zhang, J. Jung, G. A. Fiete, Q. Niu, and A.H. MacDonald, Phys. Rev. Lett. {\bf 106}, 156801 (2011).

\bibitem{BerryRMP} D. Xiao, M. C. Chang, and Q. Niu, Rev. Mod. Phys. {\bf 82}, 1959 (2010).

\bibitem{LiMorpurgo} J. Li, A. F. Morpurgo, M. B\"{u}ttiker, and I. Martin, Phys. Rev. B {\bf 82}, 245404 (2010).

\bibitem{NatPhys} J. Li, I. Martin, M. Buttiker, and A. F. Morpurgo, Nat. Phys. {\bf 7}, 38 (2011).

\bibitem{Qiao} Z. Qiao, J. Jung, Q. Niu, and A. H. MacDonald, Nano Lett. {\bf 11}, 3453 (2011).

\bibitem{Jung} J. Jung, F. Zhang, Z. Qiao, and A. H. MacDonald, Phys. Rev. B {\bf 84}, 075418 (2011).

\bibitem{Kim} A. Vaezi, Y. Liang, D. H. Ngai, L. Yang, and Eun-Ah Kim, arXiv:1301.1690 (2013).


%
%
%
%

\end{thebibliography}
\end{document}